\documentclass[twocolumn,english,groupedaddress, superscriptaddress]{revtex4-1}
\usepackage[T1]{fontenc}
\usepackage[latin9]{inputenc}
\usepackage{color}
\usepackage{textcomp}
\usepackage{amsmath}
\usepackage{graphicx}
\usepackage{babel}
\begin{document}

\title{Non-Gaussian Power Grid Frequency Fluctuations Characterized by L\'evy-stable
Laws and Superstatistics}

\author{Benjamin Sch\"afer}

\affiliation{Chair for Network Dynamics, Center for Advancing Electronics Dresden
(cfaed) and Institute for Theoretical Physics, Technical University
of Dresden, 01062 Dresden, Germany}

\affiliation{Network Dynamics, Max Planck Institute for Dynamics and Self-Organization
(MPIDS), 37077 G\"ottingen, Germany}

\author{Christian Beck}

\affiliation{Queen Mary University of London, School of Mathematical Sciences,
Mile End Road, London E1 4NS, UK}

\author{Kazuyuki Aihara}

\affiliation{Institute of Industrial Science, The University of Tokyo, Meguro-ku,
Tokyo, Japan}

\author{Dirk Witthaut}

\thanks{Contributed equally}

\affiliation{Forschungszentrum J\"ulich, Institute for Energy and Climate Research
- Systems Analysis and Technology Evaluation (IEK-STE), 52428 J\"ulich,
Germany}

\affiliation{Institute for Theoretical Physics, University of Cologne, 50937 K\"oln,
Germany}

\author{Marc Timme}

\thanks{Contributed equally}

\affiliation{Chair for Network Dynamics, Center for Advancing Electronics Dresden
(cfaed) and Institute for Theoretical Physics, Technical University
of Dresden, 01062 Dresden, Germany}

\affiliation{Network Dynamics, Max Planck Institute for Dynamics and Self-Organization
(MPIDS), 37077 G\"ottingen, Germany}
\begin{abstract}
Multiple types of fluctuations impact the collective dynamics of power
grids and thus challenge their robust operation. Fluctuations result
from processes as different as dynamically changing demands, energy
trading, and an increasing share of renewable power feed-in. Here
we analyze principles underlying the dynamics and statistics of power
grid frequency fluctuations. Considering frequency time series for
a range of power grids, including grids in North America, Japan and
Europe, we find a substantial deviation from Gaussianity best described
as L\'evy-stable and q-Gaussian distributions. We present a coarse framework
to analytically characterize the impact of arbitrary noise distributions
as well as a superstatistical approach which systematically interprets
heavy tails and skewed distributions. We identify energy trading as
a substantial contribution to today's frequency fluctuations and effective
damping of the grid as a controlling factor enabling reduction
of fluctuation risks, with enhanced effects for small power grids.
\end{abstract}
\maketitle
The Paris conference 2015 set a path to limit climate change {\emph{well below 2\textdegree C}}
\cite{The21stConferenceofthePartiestotheUnitedNationsFramework2015}.
To reach this goal, integrating renewable and sustainable energy sources
into the electrical power grid is essential \cite{ClimateChange2014}.
Wind and solar power are the most promising contributors to reach
a sustainable energy supply \cite{Jacobson2011,Schaefer2015}, but
their integration into the existing electric power system remains
an enormous challenge \cite{Turner1999,Boyle2004,Ueckerdt2015}. In
particular, their power generation varies on all time scales from
several days \cite{Heide2010} to less than a second \cite{Milan2013},
displaying highly non-Gaussian fluctuations \cite{Peinke2004}. This variability must be balanced by storage facilities and back-up plants, requiring precise control of the electric power grid.

The central observable in power grid monitoring, operation and control
is the grid frequency $f$ \cite{Machowski2011}. In case of an excess
demand, kinetic energy of large synchronous generators is converted
to electric energy, thereby decreasing the frequency. Dedicated power
plants measure this decrease and increase their generation to stabilize
the grid frequency within seconds to minutes (primary control) \cite{Machowski2011,Kundur1994}.
On longer time scales, additional power plants are activated to restore
the nominal grid frequency (secondary control). The increase of renewable
generation challenges this central control paradigm as generation
becomes more volatile and the spinning reserve decreases \cite{Ulbig2014}.
How to provide additional effective/virtual inertia is under heavy
development \cite{Delille2012,Doherty2010}. In addition, fluctuating
demand \cite{Wood2012} and fixed trading intervals \cite{NationalAcademiesofSciences2016}
already contribute to frequency deviations.

A detailed understanding of the fluctuations of power grid frequency
essentially underlies the design of effective control strategies for
future grids. Many studies for simplicity assume Gaussian noise \cite{Wood2012,Jin2005,Zhang2010,Schaefer2017,Fang2012},
while non-Gaussian effects are only rarely studied \cite{Kashima2015,Muhlpfordt2016,Anvari2016,Schmietendorf2016}.
Gaussian approaches neglect the possibility of heavy tails in the
frequency distributions and thus strong deviations from the reference
frequency posing serious contingencies particularly relevant for security
assessment. Even in studies considering non-Gaussian effects, the
connection to real data is missing \cite{Kashima2015}, realistic
but isolated wind and solar data are only numerically evaluated \cite{Anvari2016,Schmietendorf2016}
or the focus is on static power dispatch \cite{Muhlpfordt2016,Wood2012,Fang2012}
as opposed to real-time dynamics. 

It is crucial to understand how collective grid dynamics are driven
by the fluctuations originating from varying power demands, fluctuating
input generation and trading. While realistic models describing the
actual noise input of wind and solar power exist \cite{Anvari2016,Schmietendorf2016},
the impact of fluctuations on grid dynamics has been studied for selected
specific scenarios, regions or technologies only \cite{Li2013,Lauby2014}.
Furthermore, a systematic quantitative comparison of differently sized
synchronous regions based on their frequency fluctuations is needed.
It is important to forecast fluctuation statistics in grids of any
size, especially when setting up small isolated systems, e.g., on
islands or disconnected microgrids \cite{Lasseter2004}.

\begin{figure*}
\begin{centering}
\includegraphics[width=1.9\columnwidth]{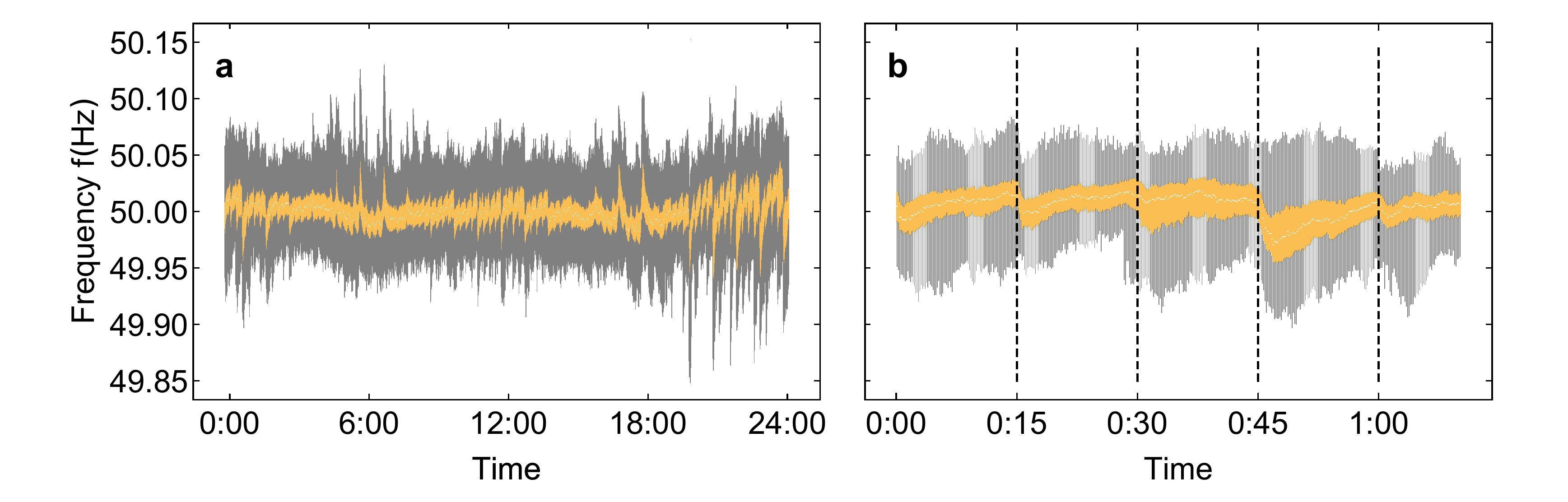}
\par\end{centering}
\caption{\textbf{Fluctuations in frequency around the reference frequency }\textbf{of
50Hz} \textbf{a}: Box plot of the 2015 data by \emph{R\'eseau
de Transport d'Electricit\'e (RTE)} \cite{RTE-UCTE2016} describing
the Continental European power grid. \textbf{b}: Zoom-in on the first
70 minutes of the frequency measurements, exposing substantial changes
in average and variance of frequencies at 15 minutes trading intervals
(indicated by the dashed lines) \label{fig:Frequency BoxWhisker}.
Each box contains data of one year for the same time instance (averaged
per minute in a). The yellow bars contain the 25\% and 75\% quartile,
the gray bars are the whiskers giving the maximum and minimum values{}
and the white line is the median value.}
\end{figure*}

In this Article, we analyze the frequency fluctuations observed in
several electric power grids from three continents. We determine and
characterize the non-Gaussian nature of these fluctuations existing
across grids in both the 60Hz and 50Hz operation regimes. Furthermore,
we propose an analytically accessible model successfully describing
these data in one consistent framework by systematically incorporating
the non-Gaussian nature of fluctuations and verify its predictions.
The analysis yields trading as a key factor for non-Gaussianity. Extracting
the effective damping for different synchronous regions via autocorrelation
measures, our work highlights that the effective grid damping as well
as the size of the grid itself serve as controlling factors to make
grid dynamics more robust. Finally, we demonstrate how superstatistics
explains heavy tails and skewness using superimposed Gaussian distributions.

\section*{Observing the statistics of Frequency Fluctuations}

The bulk frequency of a power grid fluctuates around its nominal frequency
of 60 Hz (most parts of America, western
Japan, Korea, Philippines) or 50 Hz (eastern Japan
and other countries). To understand and quantify these fluctuations,
we analyze data sets for the power grid frequency of the European Network of Transmission System Operators for Electricity (ENTSO-E) Continental
European (CE) \cite{50Hertz-UCTE2016,RTE-UCTE2016}, the Nordic \cite{Fingrid2015-2016},
Mallorcan \cite{Mallorca2015} and Great Britain (GB) \cite{UK-Frequency2016}
grids, the 50 Hz and 60 Hz regions of Japan \cite{OCCTO-Frequency2016}
as well as the Eastern Interconnection (EI) in North America \cite{US_Frequency_data},
see Supplementary Note 1 for more detailed data breakdown. The data
consist of power grid frequency measurements at one location in the
given region (two for Continental Europe) at a sampling rate between
ten measurements per second and one measurement per five minutes. 

At first glance, a typical recording of the grid frequency (Fig.~\ref{fig:Frequency BoxWhisker})
reveals that it coincides extremely well with the nominal grid reference
frequency, highlighting the efficiency of today's frequency control.
Only rarely do we observe large deviations from the nominal frequency.
These large disturbances often occur when a new power dispatch has
been settled on by trading (every 15 minutes), introducing jumps and
fluctuations of the frequency. The total variance of the frequency
fluctuations in a given region thereby depends on the size of the
grid \textendash{} larger grids are more inertial and thus tend to
have a smaller variance. 

\begin{figure}[t]
\begin{centering}
\includegraphics[width=0.95\columnwidth]{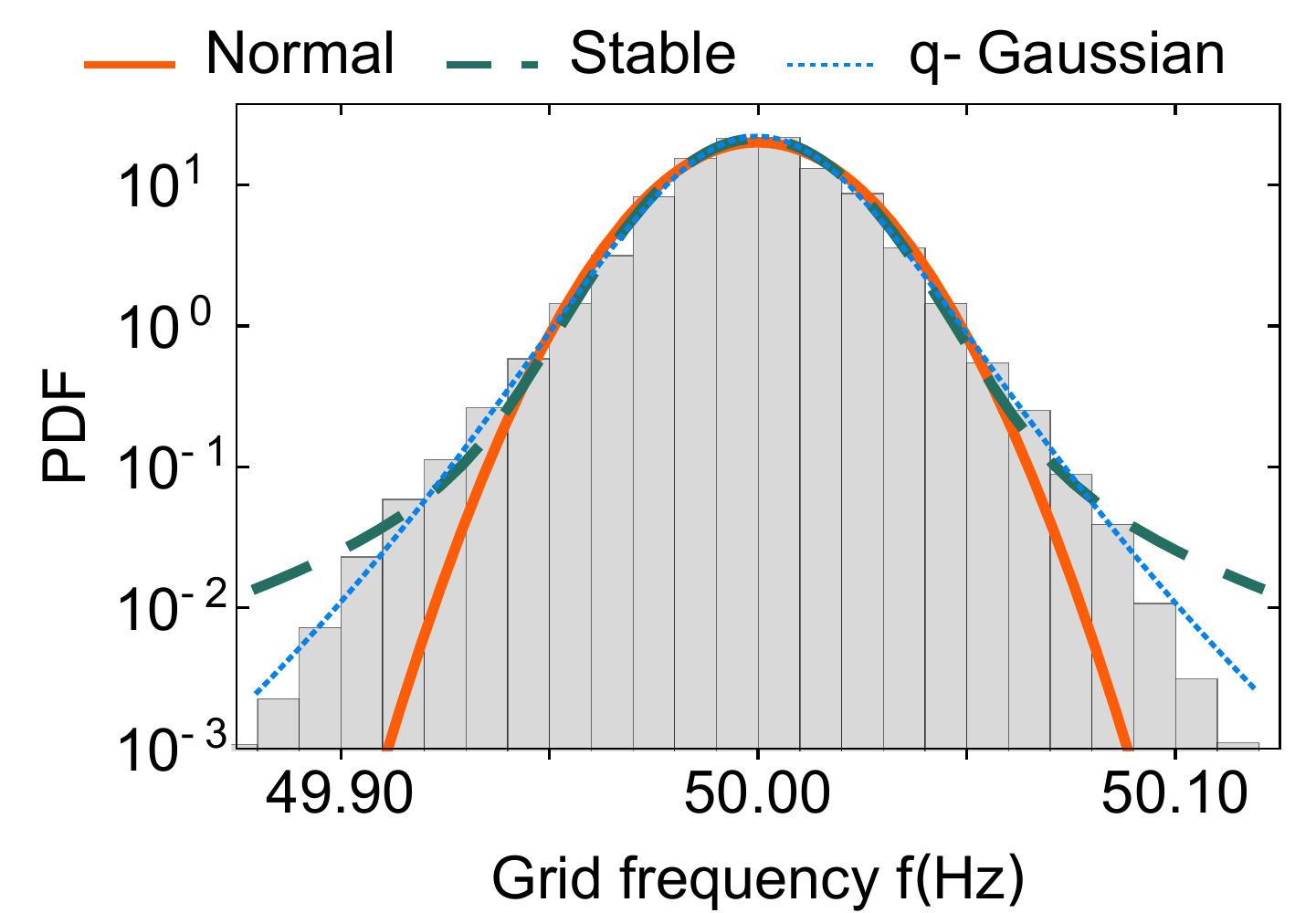}
\par\end{centering}
\caption{\textbf{Non-Gaussian nature of the frequency distribution.} 
The 2015 dataset by 50Hertz describing the CE power grid, where fitted normal, stable and q-Gaussian distributions are compared with the histogram data using a log scale for the probability density function (PDF). Deviations from a normal distribution become evident in the tails, which are more pronounced than expected for a normal distribution. 
The stability parameter of the stable distribution
is $\alpha_{S}=1.898\pm0.002$ and the deformation parameter of the
q-Gaussian distribution is $q=1.20\pm0.01$, whereas ($\alpha_{S}^{\text{Gauss}}$=2
and $q^{\text{Gauss}}$=1 for Gaussian distributions). The Lévy-stable distribution uses four fitting parameters, while the q-Gaussian uses three and the normal distribution uses two parameters.
\label{fig: Frequency  hisotgram with PDFs}}
\end{figure}

All distributions deviate from Gaussian distributions, which becomes
evident when observing their tails (Fig.~\ref{fig: Frequency  hisotgram with PDFs}).
For the Continental European, Nordic, Mallorcan and Japanese grids
large deviations from the nominal frequency are more frequent than
for a Gaussian distribution of given variance, leading to heavy tails,
as quantified, for instance, by an excess kurtosis, see Methods. The
grids of Great Britain and the Eastern Interconnection however, are
substantially skewed, i.e., they are asymmetric around the reference
frequency so that deviations towards lower frequencies are more likely
than to higher ones. 

L\'evy-stable \cite{Samorodnitsky1994} and q-Gaussian distributions
\cite{Tsallis2009} are the best fitting distributions among all distributions
tested, as identified by a maximum likelihood analysis, see Fig.~\ref{fig: Frequency  hisotgram with PDFs}
and Supplementary Note 1. Both distributions generalize a Gaussian
distribution to include heavy tails and point to two different microscopic
mechanisms underlying the frequency dynamics: q-Gaussians arise when
the power fluctuations are Gaussian on short time scales, but with
a variance or mean changing on longer time scales. In contrast, L\'evy-stable
distributions arise when the underlying power fluctuations are heavy-tailed
or skewed itself. We investigate both settings in more detail below.
\begin{figure}
\begin{centering}
\includegraphics[width=1\columnwidth]{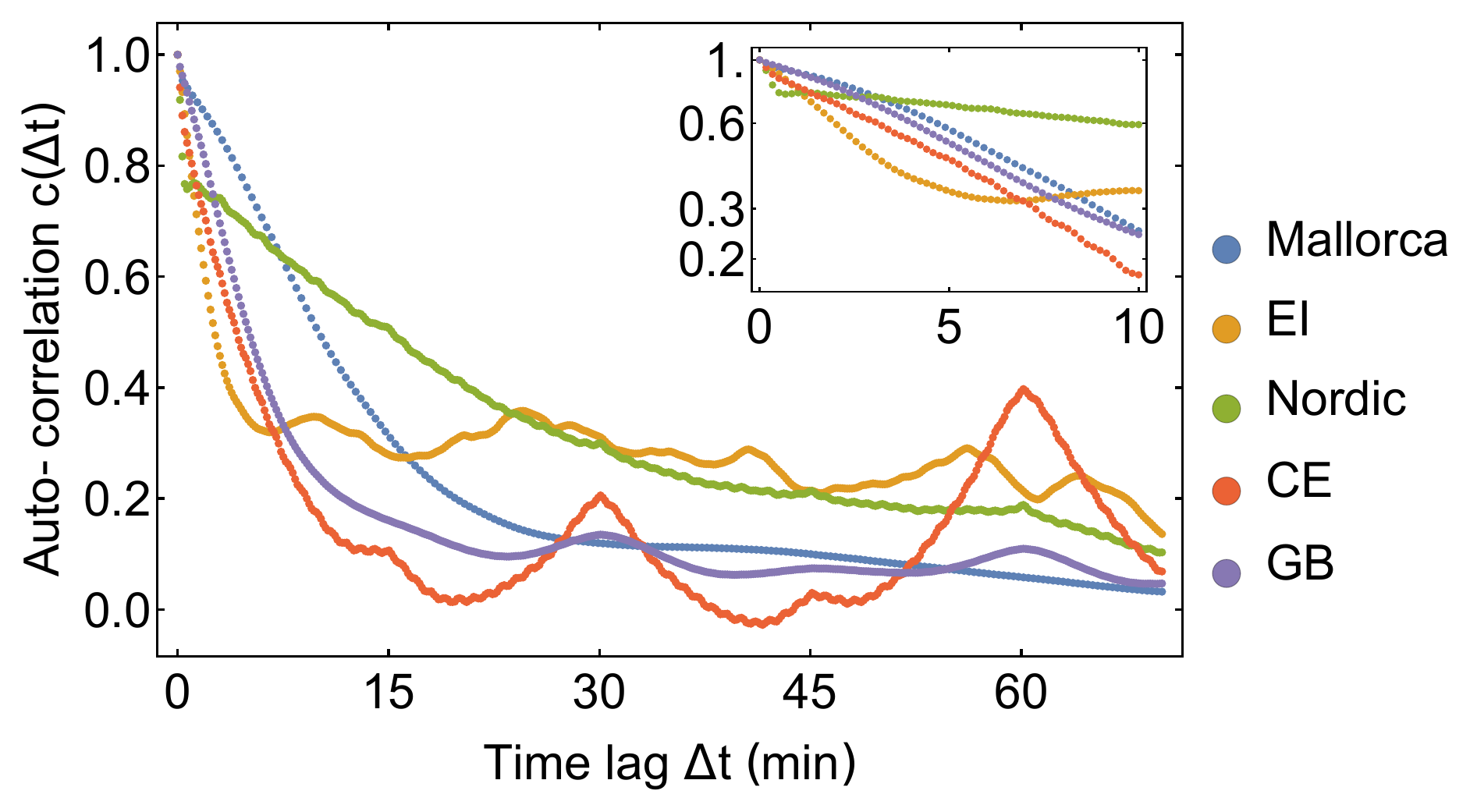}
\par\end{centering}
\caption{\textbf{Decay of the autocorrelation of the frequency} \textbf{dynamics.
}Plotted are autocorrelation measures as a function of time lag $\Delta t$
for the \emph{50Hertz} data set for Central Europe (CE) of 2015, the
Great Britain grid (GB) of 2015, the Eastern Interconnection (EI)
data for 1 day of 2015, the Nordic grid data of 2015 and Mallorcan
data of 2015. After an initial decay of the autocorrelation, peaks
emerge every 15 minutes due to trading intervals, especially pronounced
for the GB and CE grids, consistent with Fig.~\ref{fig:Frequency BoxWhisker}.
Using a log-plot in the inset allows to extract the damping of the
grid based on the assumption of exponential decay, Equation (\ref{eq:Ornstein-Uhlenbeck Autocorrelation}).
Note that the CE, GB and EI grids all display similar decay during
the initial 5 minutes. In contrast, the Nordic grid displays a fast
decay and then a slower one. The plot uses one full year of frequency
data with one second resolution for each region to generate the autocorrelation
plots. Especially the trading peaks are typically not visible when
only 24h of recordings are considered (as for the EI grid). \label{fig:50Hertz autocorrelation plots}}
\end{figure}

In addition to the aggregated data, we investigate the autocorrelation
of the recorded trajectories, extracting important events and the
characteristic time scales during which the system de-correlates.
Analyzing the autocorrelation for the Continental European grid reveals
pronounced correlation peaks every 15 minutes and especially every
30 and 60 minutes, see Fig.~\ref{fig:50Hertz autocorrelation plots}.
These regular correlation peaks appear in many grids (CE, GB, Nordic)
and are explained by the trading intervals in most electricity markets
\cite{NationalAcademiesofSciences2016}, which are often 30 or 15
minutes. Furthermore, this is also in line with the observation of
large deviations in the frequency trajectories, see Fig.~\ref{fig:Frequency BoxWhisker},
so that trading has an important impact on frequency stability. At
the beginning of a new trading interval, the production changes nearly
instantaneously and the complex dynamical power grid system needs
some time to relax to its new operational state.

The decay of the autocorrelation provides further information about
the underlying stochastic process. 
For the first minutes of each trajectory, we observe an exponential decay of the autocorrelation $c$ as a function of the time lag $\Delta t$:
\begin{equation}
c\left(\Delta t\right)\sim\exp\left(-\Delta t/\tau\right),\label{eq:Ornstein-Uhlenbeck Autocorrelation}
\end{equation}
with a typical correlation time $\tau$, as expected for elementary
stochastic processes without memory such as the Ornstein-Uhlenbeck
process \cite{Gardiner1985}. 
\begin{figure}
\begin{centering}
\includegraphics[width=0.95\columnwidth]{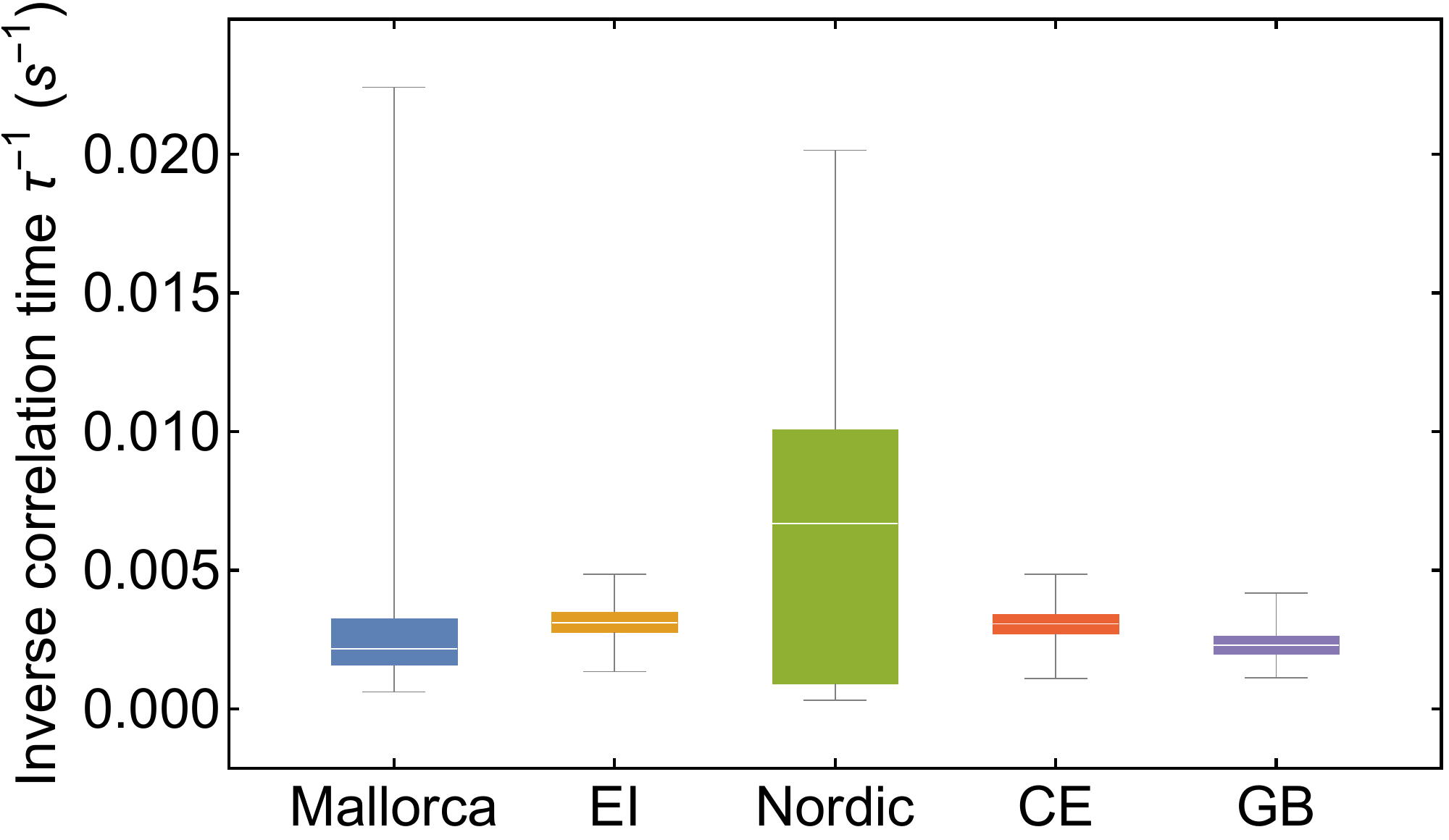}
\par\end{centering}
\caption{\textbf{Inverse correlation time of different regions.} The box plots
display the estimate of the inverse correlation time $\tau^{-1}$
based on the autocorrelation decay fitted by an exponential function,
see Equation (\ref{eq:Ornstein-Uhlenbeck Autocorrelation}). The data
are obtained by evaluating individual days of all years available
and splitting the one day of EI into 10 minute trajectories.  The
box covers the 25\% and 75\% quartile with the white line being the
median while the whiskers give the maximum and minimum values. \label{fig:Damping estimates}}
\end{figure}

We extract the inverse correlation time $\tau^{-1}$ for each available
data set and obtain values within the same order of magnitude across
all grids, see Fig.~\ref{fig:Damping estimates}. The Japanese data
set only has measurements every five minutes, hence we refrain from
estimating an autocorrelation. The inverse correlation time can be
seen as the effective damping $\gamma$ in a synchronous region with
$\gamma:=\tau^{-1}$, see below. With this in mind, it is not surprising
that all grids return values for $\gamma$ of the same order of magnitude
because the synchronous machines in these regions do not differ substantially.
This damping consists of mechanical damping, damper windings and primary
control. 

\section*{Stochastic Model of Power Fluctuations}

The variations of the grid frequency are driven by fluctuations of
power generation and demand. To link the evolution of the grid frequency
with the power injections, we make use of the well-established \emph{swing
equation} \cite{Filatrella2008,Rohden2012,Doerfler2013,Motter2013,Manik2014,Kundur1994,Machowski2011,Dewenter2015}.
Aggregating over the grid, we obtain a Fokker-Planck equation that
models the observed frequency fluctuations and allows an analytical
description of power grid frequency fluctuations.

We analyze frequency dynamics of a power grid on coarse scales. Every
node in the grid corresponds to a large generator (power plant) or
a coherent subgroup and is characterized by the phase $\theta_{i}$
and the angular velocity $\omega_{i}=2\pi\left(f_{i}-f_{R}\right)$.
Here $f_{i}$ denotes the frequency of the nodes $i=1\ldots N$ and
$f_{R}=50$Hz or $f_{R}=60$Hz, respectively, is the reference frequency
at the grid. The equations of motion of the phase and velocity are
then given by

\begin{eqnarray}
\frac{\text{d}}{\text{d}t}\theta_{i} & = & \omega_{i},\label{eq:Equation of motion power grid}\\
M_{i}\frac{\text{d}}{\text{d}t}\omega_{i} & = & P_{i}+\epsilon_{i}\xi_{i}-D_{i}\omega_{i}+\sum_{j=1}^{N}K_{ij}\sin\left(\theta_{j}-\theta_{i}\right),\nonumber 
\end{eqnarray}
where we have at each node $i$: inertia $M_{i}$, voltage phase angle
$\theta_{i}$, mechanical power $P_{i}$, random noise $\xi_{i}$
with noise amplitude $\epsilon_{i}$, damping $D_{i}$ and the coupling
matrix $K_{ij}$ which is determined by the transmission grid topology.
The operating state of a power grid is characterized by a stable fixed
point of the swing equation (\ref{eq:Equation of motion power grid}).
The fixed point fulfills $\omega_{i}^{*}=0$ which is equivalent to
all machines working at the reference frequency $f_{R}=50$Hz or $f_{R}=60$Hz.
At the stable operation point the frequencies at all nodes are equal:
$\omega_{i}=\bar{\omega}$. Deviations are only observed during system-wide
failures or transiently after serious contingencies or major topology
changes \cite{Machowski2011,Kundur1994}. To obtain the effective
equation of motion of the bulk angular velocity $\bar{\omega}$, we
assume a homogeneous ratio of damping and inertia throughout the network,
$\gamma=D_{i}/M_{i}$ \cite{Weixelbraun2009} as well as symmetric
coupling $K_{ij}=K_{ji}$ and assume that the power is balanced $\sum_{i=1}^{N}P_{i}=0$
\emph{on average} \cite{Manik2014}. Setting $M:=\sum_{i}M_{i}$,
the dynamics of the bulk angular velocity $\bar{\omega}:=\sum_{i=1}^{N}M_{i}\omega_{i}/M$
is governed by the \emph{Aggregated Swing Equation} (see also \cite{Ulbig2014})
\begin{equation}
\frac{\text{d}}{\text{d}t}\bar{\omega}=-\gamma\bar{\omega}+\bar{\epsilon}\bar{\xi}\left(t\right).\label{eq:Stochastic equation of motion}
\end{equation}
This aggregated swing equation no longer requires precise knowledge
of the parameters of a given region but depends on the effective damping
$\gamma$, the aggregated noise amplitude $\bar{\epsilon}$ and the
statistics of the random noise $\bar{\xi}$, all characterizing the
overall frequency dynamics, see Methods and Supplementary Note 2 for
details. We note that the damping $\gamma$ integrates contributions
from damper windings and primary control actions alike. Finally, both
damping $\gamma$ and the noise amplitude $\bar{\epsilon}$ could
easily change over time, e.g., due connection of certain grids or
day/night differences. We cover this scenario in the section on superstatistics.

The bulk angular velocity $\bar{\omega}$ (and thereby the grid frequency)
is not following a deterministic evolution but is influenced by stochastic
effects, given by the aggregated power fluctuations $\bar{\xi}$.
Hence, we characterize a given grid by the probability distribution
function (PDF) of the bulk angular velocity $p\left(\bar{\omega}\right)$,
similar to the frequency distribution plotted in Fig.~\ref{fig: Frequency  hisotgram with PDFs}.
A wide distribution, i.e. one with high standard deviation, or one
with heavy tails, i.e., high kurtosis, displays large deviations more
often and is thereby less stable than a narrower distribution. 

The central decision when modeling stochastic dynamics is how to describe
the noise $\xi$ which is generated from some probability distribution.
Explicit choices of noise distributions are covered here and in Supplementary
Notes 2 and 3 for Gaussian and non-Gaussian noise, respectively and
extended to noise drawn from a Gamma distribution \cite{Carpaneto2008,T.Soubdhan2009}
in Supplementary Note 4. Given the distribution of $\xi$, we then
formulate and solve a Fokker-Planck equation \cite{Gardiner1985}
to obtain an analytical description of the distribution of $\bar{\omega}$.

The simplest noise model assumes the noise $\xi_{i}$ as independent
Gaussian noise based on the often-used central limit theorem. It states
that the sum of independent random numbers drawn from any fixed distribution
with finite variance approaches a Gaussian distribution if the sample
is sufficiently large \cite{Gardiner1985}. In our setting, the sum
consists of all contributions to the noise by consumers, renewables,
trading etc. The Fokker-Planck equation describing the time-dependent
probability density function $p\left(\bar{\omega},t\right)$ follows
then as 
\begin{equation}
\frac{\partial p}{\partial t}=\gamma\frac{\partial}{\partial\bar{\omega}}\left(\bar{\omega}p\right)+\frac{1}{2}\sum_{i=1}^{N}\frac{\epsilon_{i}^{2}}{M^{2}}\frac{\partial^{2}p}{\partial\bar{\omega}^{2}},\label{eq:Fokker-Planck mean frequency}
\end{equation}
which is the well-known Ornstein-Uhlenbeck process \cite{Gardiner1985}.
The stationary distribution

\begin{equation}
p\left(\bar{\omega}\right)=\sqrt{\frac{\gamma M^{2}}{\pi\sum_{i=1}^{N}\epsilon_{i}^{2}}}\exp\left[-\bar{\omega}^{2}\frac{\gamma M^{2}}{\sum_{i=1}^{N}\epsilon_{i}^{2}}\right],\label{eq:(ordinary) Fokker-Planck solution}
\end{equation}
of (\ref{eq:Fokker-Planck mean frequency}) characterizes the steady
state of the grid as mathematically defined by $\partial p/\partial t$=0,
see \cite{Gardiner1985} as well as Methods and Supplementary Notes
2 and 6 for details. 

Crucially, Equation (\ref{eq:(ordinary) Fokker-Planck solution})
is again a Gaussian distribution of $p\left(\bar{\omega}\right)$,
i.e., a Gaussian distribution for the power feed-in fluctuations results
in a Gaussian frequency distribution. Assuming we know the damping
$\gamma$, noise amplitudes $\epsilon_{i}$ and the total inertia
$M$, we are able to compute the expected frequency distribution analytically.
Furthermore, the Ornstein-Uhlenbeck autocorrelation exactly follows
an exponential decay with characteristic time determined by the damping
$\tau=1/\gamma$.

Under which conditions do we need to include non-Gaussian effects
in the stochastic modeling? When applying the central limit theorem,
one explicitly assumes \emph{finite} variance. However, solar and
wind fluctuations are known to display heavy tails
\cite{Anvari2016,Milan2013} and contribute to the fluctuations in
the power grid. Hence, to describe deviations from normal distributions,
including heavy tails and skewed distributions, we need to base the
input noise $\xi$ on a non-Gaussian noise generating process \cite{Denisov2009}.
This requires generalized Fokker-Planck equations, see Supplementary
Note 3. These generalized equations characterize fluctuations based
on noise input distributed according to, e.g., a L\'evy-stable law.
These L\'evy-stable distributions include heavy tails and skewed distributions,
as often observed in nature \cite{Peinke2004} and are a reasonable
fit for the frequency data, see Fig.~\ref{fig: Frequency  hisotgram with PDFs}.
Stable distributions are characterized by a stability parameter $\alpha_{S}\in(0,2]$,
which determines the heavy tails, a skewness parameter $\beta_{S}$
and a scale parameter $\sigma_{S}$, which is similar to the standard
deviation for Gaussian distributions \cite{Samorodnitsky1994}.

Inputting power fluctuations $\xi$ drawn from a stable distribution
into the stochastic Equation (\ref{eq:Stochastic equation of motion})
also results in grid frequency fluctuations characterized by a stable
distribution, considered as the 'output' of Equation (\ref{eq:Stochastic equation of motion}).
Between input and output distributions, only the scale parameter is
modified whereas the skewness $\beta_{S}$ (asymmetry) and the stability
parameter $\alpha_{S}$ (heavy-tail-ness) are preserved. In particular,
the scale parameter $\sigma_{S}^{\text{in}}$ of the input distribution
changes to that of the output distribution $\sigma_{S}^{\text{out}}$
following the map (Supplementary Notes 3 and 6)
\begin{equation}
\sigma_{S}^{\text{in}}=\frac{1}{\sqrt{2}M}\left[\sum_{i=1}^{N}\epsilon_{i}^{\alpha_{S}}\right]^{1/\alpha_{S}}\mapsto\,\sigma_{S}^{\text{out}}=\frac{\sigma_{S}^{\text{in}}}{\left(\gamma\alpha_{S}\right)^{1/\alpha_{S}}}.\label{eq:transformed scale parameter}
\end{equation}
 We emphasize this remarkable and unique property of stable distributions
\cite{Samorodnitsky1994} for linear models: If the input power fluctuations
are distributed according to a stable distribution, the output frequency
fluctuations are distributed according to the same family of distributions,
with only one parameter transformed. This property holds for any linear
stochastic process, including the aggregated swing equation (\ref{eq:Stochastic equation of motion}).
The same happens for Gaussian distributions since they constitute
a subclass of stable distributions in the limit $\alpha_{S}\rightarrow2$.
These properties are in stark contrast to those of non-stable distributions,
see Supplementary Notes 3 and 4. 

What are the consequences of relation (\ref{eq:transformed scale parameter})?
Making the output frequency distribution narrower, i.e., reducing
risks of extreme events, requires $\sigma_{S}^{\text{out}}$ to be
as small as possible. However, increasing the share of renewables
by rebuilding the energy system is expected to increase the noise
amplitudes $\epsilon_{i}$. In addition, trading impacts the frequency
fluctuations and thereby also contributes to the noise amplitudes
(Fig.~\ref{fig:Frequency BoxWhisker}). However, fluctuations are
efficiently reduced by increasing the effective damping $\gamma$
or the inertia $M$, see Eq. (\ref{eq:transformed scale parameter}).

With the previous results, we are able to quantify the intuitive statement
that larger regions have more inertia and hence narrower distributions
by explicitly comparing the scale parameters (proportional to standard
deviations in the case of $\alpha_{S}=2$) of two different regions
as follows: 
\begin{equation}
\sigma_{S\,2}^{\text{out}}=\sigma_{S\,1}^{\text{out}}\frac{m_{1}}{m_{2}}\left(\frac{\gamma_{1}N_{1}^{\alpha_{S}-1}}{\gamma_{2}N_{2}^{\alpha_{S}-1}}\right)^{1/\alpha_{S}},\label{eq:Scale parameter comparison of two regions}
\end{equation}
assuming identical stability parameters $\alpha_{S}$ and average
inertia $m_{\mu}=M_{\mu}/N_{\mu},\,\mu\in\left\{ 1,2\right\} $. Equation
(\ref{eq:Scale parameter comparison of two regions}) shows that a
smaller region ($N_{2}<N_{1}$) needs larger damping than a larger
region ($\gamma_{2}\overset{!}{>}\gamma_{1}$) or has a broader distribution
with $\sigma_{S\,2}^{\text{out}}>\sigma_{S\,1}^{\text{out}}$, i.e.,
a higher risk of large deviations from the stable operational range.
The scaling is given by the scale parameter $\sigma_{S}\sim N^{\left(\alpha_{S}-1\right)/\alpha_{S}}$,
where the simple square root law is recovered only in the case of
Gaussian distributions $\left(\alpha_{s}=2\right)$. Also, it reveals
that decreasing inertia proportionally increases the scale parameter.

\begin{figure}[t]
\begin{centering}
\includegraphics[width=0.95\columnwidth]{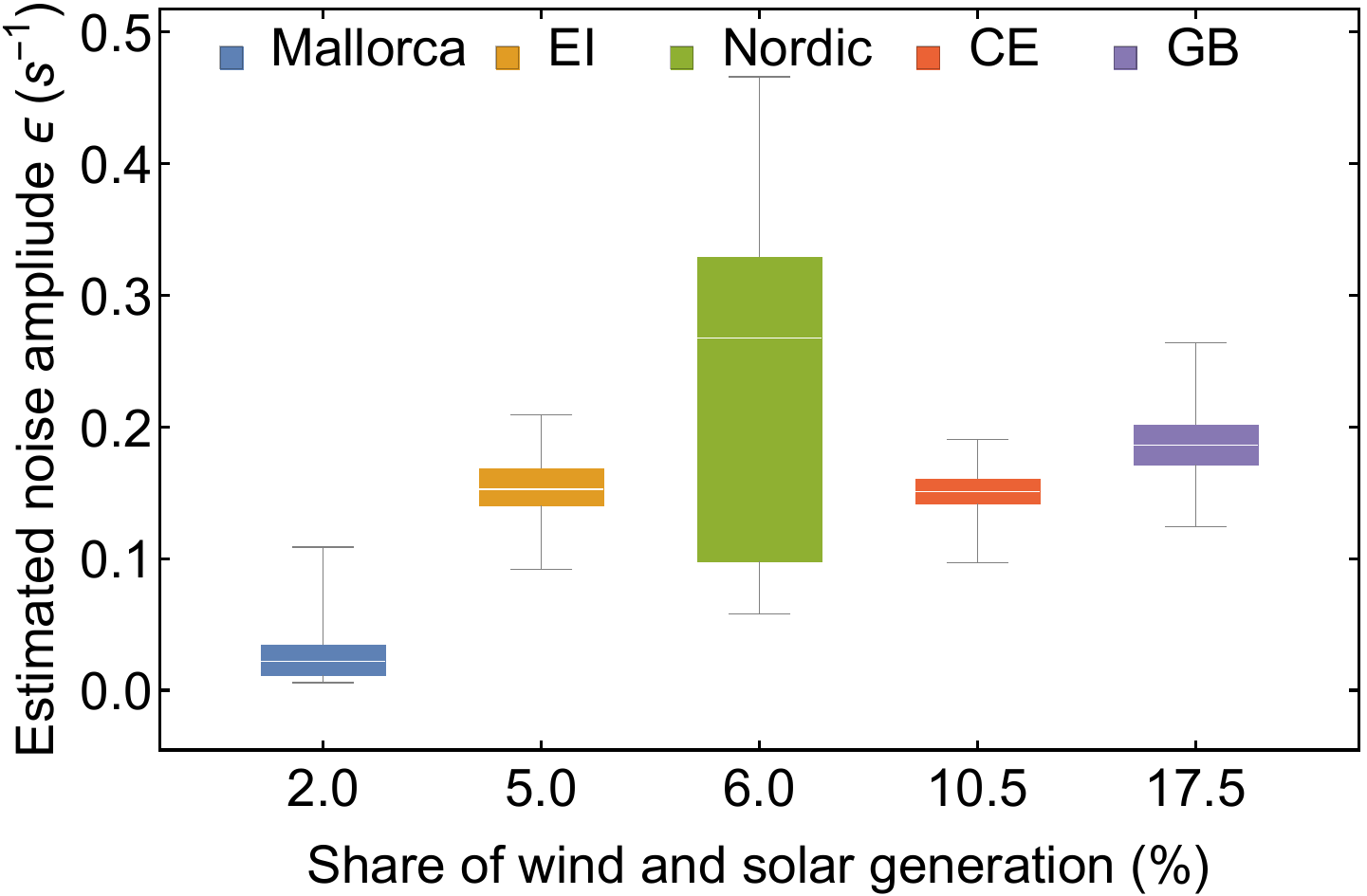}
\par\end{centering}
\caption{\textbf{Noise amplitudes for European and American grids.} The noise
amplitude tends to increase with the shares of intermittent renewables.
The noise amplitude $\epsilon$ for each grid is calculated assuming
that it is identical at each node $\epsilon_{i}=\epsilon$ and assuming
homogeneous inertia. The power production is normalized with respect
to the Eastern Interconnection (EI) generation for the ENTSO-E grids
of Continental Europe (CE), Mallorca, Nordic and Great Britain (GB).
Frequency data of all regions and Equation (\ref{eq:Predicted noise amplitude for STABLE NOISE})
is used to compute the noise amplitude $\epsilon$ which we expect
to be similar in all regions, providing a self-consistency check of
our theory. The box plot is obtained by using different damping, standard
deviation estimates, etc. for each day of multiple years. \label{fig:Noise amplitudes for different regions}
The data for the Nordic grid has large uncertainty due to the two
different correlation time scales. See Supplementary Note 1 for details
on the data.}
\end{figure}

Furthermore, we estimate the order of magnitude of the expected noise
amplitude 
\begin{equation}
\epsilon=\sigma_{S}^{\text{out}}m\left(\alpha_{S}\gamma N^{\alpha_{S}-1}\right)^{1/\alpha_{S}}.\label{eq:Predicted noise amplitude for STABLE NOISE}
\end{equation}
by computing the scaling from (\ref{eq:transformed scale parameter})
for typical noise contributions of the order of $\epsilon_{i}=\epsilon$.
Based on pure frequency measurements, every quantity is available
for each synchronous region: We estimate the output scale parameter
$\sigma_{S}^{\text{out}}$ and stability parameter $\alpha_{S}$ from
the histogram data. We assume that the number of nodes $N$ is directly
proportional to the total electricity production of a region per year
\cite{ENTSO-EProductionData2016,EIA-411EnergyReport2016}. Since a
process driven by stable noise has no defined autocorrelation function
\cite{Samorodnitsky1994}, we approximate its autocorrelation with
the Ornstein-Uhlenbeck process and thereby derive an estimate for
the damping $\gamma$. With these estimates and Equation (\ref{eq:Predicted noise amplitude for STABLE NOISE})
we plot the noise amplitudes for different regions in Fig.~\ref{fig:Noise amplitudes for different regions}.
The estimated noise amplitude tends to increase with increasing share
of intermittent renewable generation (wind and solar) in a given region.
Nevertheless, this relationship is not very strict and frequency disturbances
at trading intervals, see Fig.~\ref{fig:Frequency BoxWhisker} demonstrate,
that at least today trading and demand fluctuations are contributing
substantially to frequency fluctuations.

\section*{Superstatistics}

Instead of modeling the underlying stochastic process as non-Gaussian,
we may interpret the observed statistic as a superposition of multiple
Gaussians, leading to \emph{superstatistics}, explaining heavy tails
and skewness \cite{Beck2003,Chechkin2017}. 

For our superstatistical approach we use Equation (\ref{eq:Stochastic equation of motion})
with Gaussian noise $\bar{\xi}$ 
\begin{equation}
\frac{\text{d}}{\text{d}t}\bar{\omega}=-\gamma\bar{\omega}+\bar{\epsilon}\bar{\xi}\left(t\right),
\end{equation}
which yields a Gaussian distribution, see Eq. (\ref{eq:(ordinary) Fokker-Planck solution}). 

\begin{figure*}
\begin{centering}
\includegraphics[width=1.9\columnwidth]{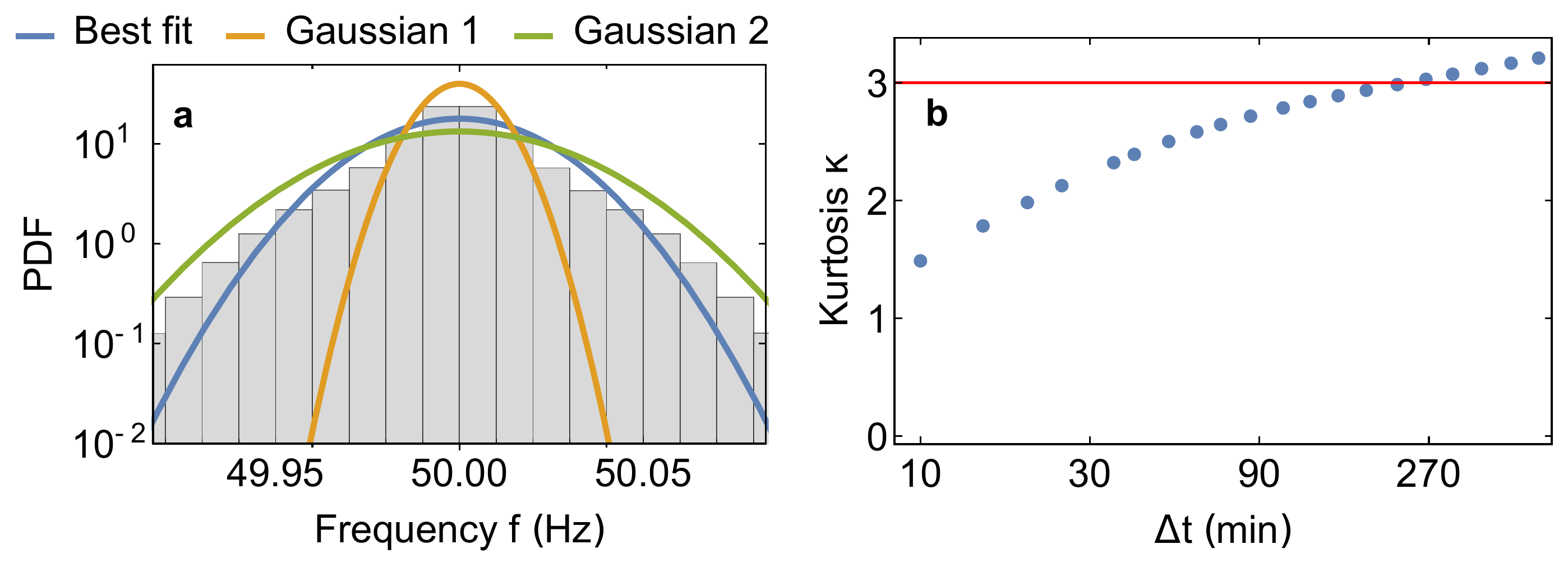}
\par\end{centering}
\caption{\textbf{Superimposed Gaussian distributions leading to heavy tails.}
\textbf{a}: When the stochastic process follows two different Gaussian
distributions (orange and green) and the data is aggregated (gray
histogram), the result is a heavy-tailed distribution
which is not Gaussian. Consequently, Gaussian fits (blue curve) tend
to underestimate its tails. Assuming such a structure for the real
frequency measurements, the frequency recordings are split into trajectories
of length $\Delta t$ each and the kurtosis is computed. \textbf{b}:
The average kurtosis of the Japanese 60Hz data set in dependence of
the length of $\Delta t$. For very small $\Delta t$ the distribution
has lighter tails than a Gaussian while using the full data set or
large $\Delta t$ leads to the earlier observed heavy tails. The long
time scale $T$, during which the distribution changes, is determined
as $\kappa\left(\Delta t=T\right)=3$. \label{fig:Excess-kurtosis}}
\end{figure*}

What changes when the damping $\gamma$ is no longer constant over
time? Both control actions and physical damping contribute to $\gamma$
and change over time when certain power plants are connected and others
are shut down. Similarly, the noise amplitude $\bar{\epsilon}$ of
the system depends on which consumers are currently active, whether
it is day or night, which renewables are connected
and more. Hence, it is appropriate to replace our static parameters
$\gamma$ and $\bar{\epsilon}$ by dynamical parameters that change
over time with a typical time scale $T$. When applying superstatistics,
we assume that the time scale $T$ is large compared to the intrinsic
time scale of the system, which is given by the autocorrelation time
scale, namely $T\gg\tau=1/\gamma$. Then, the stochastic process finds
an equilibrium with an approximately Gaussian distribution determined
by the current noise and damping. When these parameters change, the
frequency distribution becomes a Gaussian distribution with different
standard deviation. In Fig.~\ref{fig:Excess-kurtosis}a we demonstrate
how just a few Gaussian distributions with different standard deviations
give rise to an excess kurtosis and in Supplementary Note 5 we show
how two Gaussian distributions with shifted means result in a skewed
distribution. 

We extract the long time scale $T$ from the data and compare it to
the intrinsic short time scale of the system. The short time scale
$\tau=1/\gamma$ is based on the exponential decay of the autocorrelation
of the time series of $\bar{\omega}$ yielding a range of $\tau\approx200...550\,s$
for all grids. The long time scale $T$ is governed by the idea that
the superstatistical ensemble has heavier tails than a normal distribution
but that for a given typical time scale $T$ an equilibrium distribution
emerges that is approximately Gaussian. Given a time series $x\left(t\right)$
with mean $\bar{x}$, we compute the local kurtosis $\kappa\left(\Delta t\right)$
for different time intervals $\Delta t$ and choose the large time
scale $T$ by $\kappa\left(\Delta t=T\right)=3$ \cite{Beck2003}.
Similarly, we compute the time for which the average skewness is zero
to extract the long time scale for the Great Britain or Eastern Interconnection
grids, see Methods and Supplementary Note 5 for details and Fig.~\ref{fig:Excess-kurtosis}
for an example for Japan.

All synchronous regions return large but different long time scales
$T$. We determine the long time scales to be of the order of $T\approx1\ldots5\,h$
with small values in Mallorca and the Eastern Interconnections and
large values in Continental Europe and Japan, hinting to distinct
underlying mechanisms how damping and noise change in each region.
Compared to the intrinsic short time scale $\tau\sim200...550\,s$,
the long time scale $T$ is larger by at least one order of magnitude.
Hence, the superstatistical approach is justified, i.e., it is valid
to interpret the heavy tails as a result of superimposing Gaussians.

\begin{figure}[t]
\begin{centering}
\includegraphics[width=0.9\linewidth]{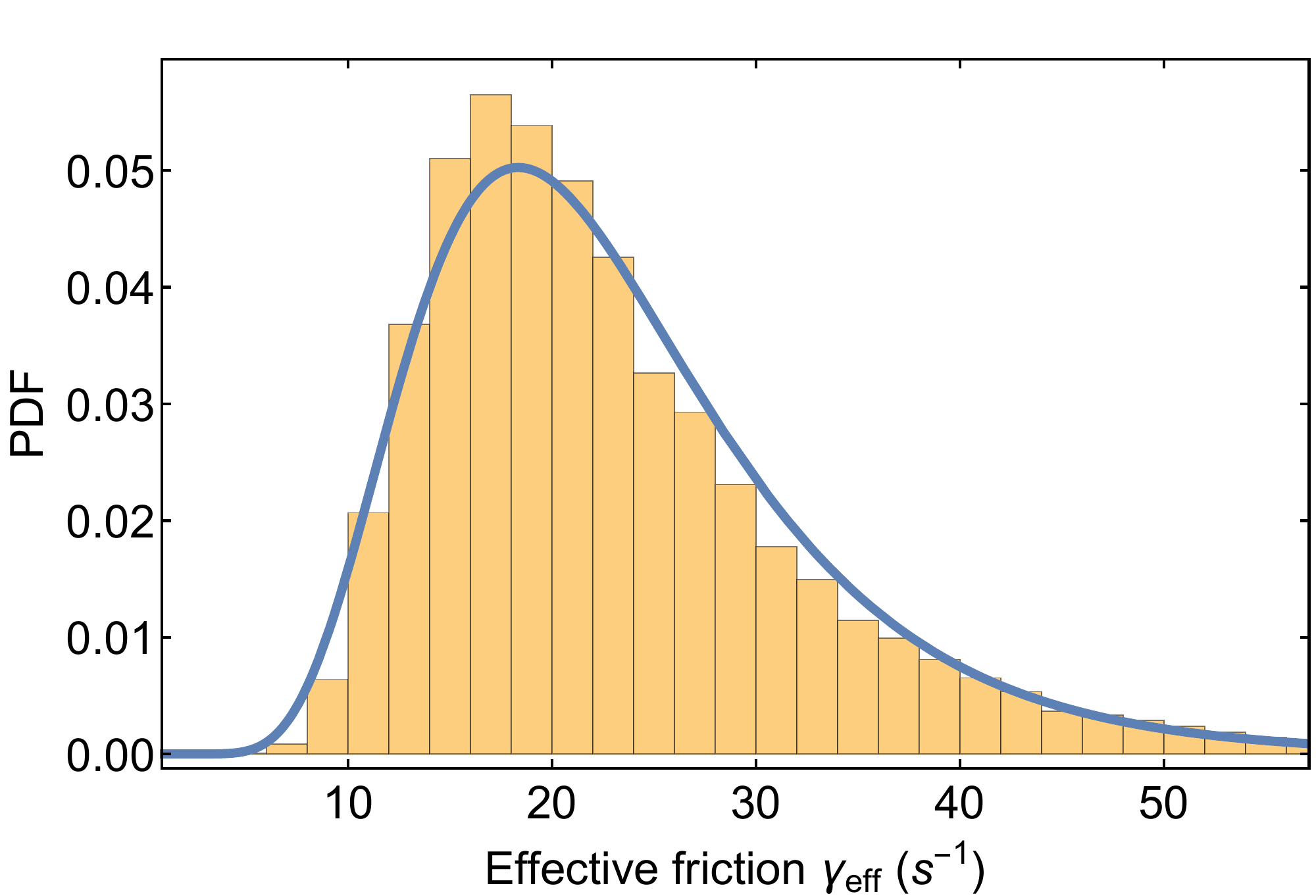}
\par\end{centering}
\caption{\textbf{Self-consistency test of superstatistics.} Plotted is the
histogram of the effective friction $\gamma_{\text{eff}}$ based on
the Japanese 60Hz frequency measurements which is well-described by
a log-normal distribution. Such a distribution of the effective friction
$\gamma_{\text{eff}}$, directly leads to q-Gaussian distributions
of the aggregated data, see Supplementary Note 5. Other data sets
are also approximated by log-normal distributions, see Supplementary
Note 5. \label{fig:Japanese beta distribution}}
\end{figure}

Finally, we perform another consistency check of the superstatistical
approach and extract the distribution of the \emph{effective friction}
$\gamma_{\text{eff}}$ \cite{Beck2003}, see Methods. Based on general
results on superstatistics, we expect the effective friction to follow
a $\chi^{2}$, inverse $\chi^{2}$ or log-normal distribution \cite{Clark1973,B.Castaing1990},
which then leads to an approximate q-Gaussian distribution of the
frequency, see Supplementary Note 5 for a derivation. In the case
of the Japanese 60Hz region the distribution of $\gamma_{\text{eff}}$
is well-described by a log-normal distribution again supporting the
superstatistical approach, see Fig.~\ref{fig:Japanese beta distribution}.

\section*{Discussion}

In summary, we have analyzed power grid frequency fluctuations by
applying analytical stochastic methods to time series of different
synchronous regions across continents including North America, Japan
and different European regions. Based on bulk frequency measurements,
we have identified trading as a substantial  source of fluctuations
(Figs.~\ref{fig:Frequency BoxWhisker} and \ref{fig:50Hertz autocorrelation plots}).
Although frequency fluctuations and power uncertainty are often modeled
as Gaussian distributions \cite{Wood2012,Jin2005,Zhang2010,Schaefer2017,Fang2012},
we pinned down and quantified substantial deviations from a Gaussian
form, including heavy tails and skewed distributions (Fig.~\ref{fig: Frequency  hisotgram with PDFs}).

Obtaining an analytical description of a complex system allows deeper
insight into it. Hence, condensing the analysis of frequency fluctuations
in power grids via a second order nonlinear dynamics, the swing equation,
and neglecting spatial correlations, we derived (generalized) Fokker-Planck
equations for the bulk angular velocity $\bar{\omega}$. We obtained
precise predictions on how power fluctuations impact the distribution
of fluctuations of the grid frequency. Furthermore, our approach identifies,
besides grid size, an increasing effective damping and inertia as
a controlling factor for reducing fluctuation-induced risks. By incorporating
smart grid control mechanisms \cite{Schaefer2015} or increasing generator
droop control \cite{Machowski2011}, modifying effective damping may
therefore reliably reduce the likelihood of large fluctuations in
power grids \cite{Schoell2008}. Finally, our analytical theory is
able to compare differently sized grids, predict
fluctuations based on the size and inertia of the grid (Equation (\ref{eq:Scale parameter comparison of two regions})).
Crucially, our mathematical framework goes beyond the simple $N^{-1/2}$
scaling of Gaussian noise.

The results offer two approaches to model power grids under uncertainty:
First, an optimization could include the non-Gaussian nature of the
distribution by incorporating non-Gaussian noise, e.g. in the form
of L\'evy-stable noise. Alternatively, we demonstrated that the distributions
are also well explained by a \emph{superstatistics} approach where
the non-Gaussian nature of the distributions arises by superimposing
different Gaussian distributions. Especially when modeling shorter
time scales of one hour or below, a Gaussian approach is supported
by our results. Studies aiming to cover time scales of full months
or years, however, have to account for changing mean and variance
of the assumed Gaussian distribution or explicitly model non-Gaussian
distributions, going beyond current Gaussian approaches \cite{Wood2012,Jin2005,Zhang2010,Schaefer2017,Fang2012}. 

The findings reported above have a number of implications for the
operation and design of current and future energy systems. First,
as trading induces large frequency fluctuations, designing new electricity
markets and limiting frequency fluctuations are highly interlinked,
especially when considering the implementation of smart grid concepts
\cite{Fang2012,Schaefer2015}. Second, knowing the temporal correlation
structure of fluctuations helps predicting increasing and decreasing
likelihoods of large amplitude events, thereby enabling mitigation
strategies to be applied on time scales that make them most efficient.
Finally, deriving the scaling of fluctuations as a function of grid
parameters, especially the grid size, should be very useful when setting
up isolated grids, e.g. microgrids with a specified frequency quality
as damping and control needs can easily be estimated by the approach
introduced above. This may also be of use for larger synchronous regions
when facing a decreasing inertia $M$.

Moreover, applying similar stochastic methods to power grids also
raises a range of additional questions: How does correlated noise
impact the frequency statistics? Does the predicted scaling of fluctuations
with the grid size hold for a larger collection of independent power
grids and in particular very small islands or microgrids? Can we disentangle
damping and primary control to explain the differences of long time
scales among different regions? These questions require further careful
data analysis in future work, involving substantially more data of
microgrids, work that could inspire further collaboration including
a range of academic fields as well as public institutions and industry.

\section*{Methods}

\subsection*{Moments of the frequency distributions}

Deviations from Gaussian distributions as observed in Fig.~\ref{fig: Frequency  hisotgram with PDFs}
are quantified in a model independent way using moments of the frequency
distribution: Given $M$ measurements of a discrete stochastic variable
$f$, e.g. the grid frequency, as $f_{1}$, $f_{2}$,...,$f_{M}$,
its $n$-th moment is defined as
\begin{equation}
\mu_{n}:=\frac{1}{M}\sum_{i=1}^{M}f_{i}^{n}.
\end{equation}
The first moment of a distribution is the mean $\mu_{1}\equiv\mu$.
Instead of the second moment, the centralized second moment, i.e.,
the variance is more commonly used. It is defined as
\begin{equation}
\sigma^{2}:=\frac{1}{M}\sum_{i=1}^{M}\left(f_{i}-\mu\right)^{2}.
\end{equation}
Finally, we use the normalized third and fourth moments, the skewness
$\beta$ and kurtosis $\kappa$, respectively, which are defined as
\begin{eqnarray}
\beta & := & \frac{1}{M}\sum_{i=1}^{M}\left(\frac{f_{i}-\mu}{\sigma}\right)^{3},\\
\kappa & := & \frac{1}{M}\sum_{i=1}^{M}\left(\frac{f_{i}-\mu}{\sigma}\right)^{4}.
\end{eqnarray}
A Gaussian distribution is symmetric and hence the skewness $\beta$
equals zero. A non-zero skewness implies a distribution that is not
symmetric around the mean but is more pronounced in one direction.
The kurtosis meanwhile quantifies the extremity of the tails. A Gaussian
distribution has $\kappa^{\text{Gauss}}=3$ while a higher value indicates
an increased likelihood of large deviations. For instance, the continental
European grid displays a kurtosis of $\kappa^{\text{CE}}=4.0\pm0.1$.

\subsection*{Normally distributed noise}

For Eq. (\ref{eq:Fokker-Planck mean frequency}) we took the sum over
multiple noise realizations that follow a normal distribution: Let
$\xi_{i}$ be random variables following a normal distribution, i.e.,

\begin{equation}
\xi_{i}\sim N\left(0,1\right),
\end{equation}
where $N\left(0,1\right)$ denotes a normal distribution with mean
$0$ and standard deviation $1$. Then, the sum of identically and
independently distributed random variables $\xi_{i}$ given as 
\begin{equation}
\bar{\epsilon}\bar{\xi}:=\sum_{i=1}^{N}\epsilon_{i}\xi_{i}
\end{equation}
 is distributed like a single normal distribution \cite{Samorodnitsky1994}
\begin{equation}
\bar{\epsilon}\bar{\xi}\sim N\left(0,\sqrt{\sum_{i=1}^{N}\epsilon_{i}^{2}}\right).
\end{equation}

\subsection*{Superstatistics}

In Figs.~\ref{fig:Excess-kurtosis} and \ref{fig:Japanese beta distribution}
we extract the local kurtosis and effective damping from the time
series as follows. Let $x\left(t\right)$ be a time series of random
measurements with a mean $\bar{x}$. To test whether $x\left(t\right)$
is aggregated by drawing from multiple distributions, we compute the
local kurtosis as: 
\begin{equation}
\kappa\left(\Delta t\right)=\frac{1}{t_{\text{max}}-\Delta t}\int_{0}^{t_{\text{max}}-\Delta t}\frac{\left\langle \left(x-\bar{x}\right)^{4}\right\rangle _{t_{0},\Delta t}}{\left\langle \left(x-\bar{x}\right)^{2}\right\rangle _{t_{0},\Delta t}^{2}}\text{d}t_{0},
\end{equation}
where $\left\langle ...\right\rangle _{t_{0},\Delta t}=\int_{t_{0}}^{t_{0}+\Delta t}...\text{d}t$.
We do so for several values of $\Delta t$ and choose $T$ so that
$\kappa\left(\Delta t=T\right)=3$, i.e., averaging over a time scale
$T$ , there is no excess kurtosis and locally the variable $x$ is
subject to Gaussian noise. 

The effective friction $\gamma_{\text{eff}}$, which is changing over
time, is then computed as
\begin{equation}
\gamma_{\text{eff}}\left(t_{0}\right)=\frac{1}{\left\langle x^{2}\right\rangle _{t_{0},T}-\left\langle x\right\rangle _{t_{0},T}^{2}}.
\end{equation}
Following \cite{Beck2003} we expect $\gamma_{\text{eff}}$ to follow
a log-normal or alternatively a $\chi^{2}$ or inverse $\chi^{2}$
distribution as those lead to q-Gaussian distributions of $x$, see
Supplementary Note 5. 

\subsection*{Data availability}

Frequency recordings are publicly available at the respective references
for the CE, GB, Nordic and Japanese regions \cite{50Hertz-UCTE2016,RTE-UCTE2016,Fingrid2015-2016,UK-Frequency2016,OCCTO-Frequency2016}.
Frequency data for Mallorca \cite{Mallorca2015}
were provided by Eder Batista Tchawou Tchuisseu. Data for the Eastern
Interconnection \cite{US_Frequency_data} were provided by Micah Till.
All data that support the results presented in the figures of this
study are available from the authors upon reasonable request.

\bibliographystyle{naturemag}

\begin{acknowledgments}
We gratefully acknowledge support from the Federal Ministry of Education
and Research (BMBF grant no. 03SF0472A-F to M.T. and D.W.), the Helmholtz
Association (via the joint initiative ``Energy System 2050 - A Contribution
of the Research Field Energy'' and the grant no.VH-NG-1025 to D.W.),
the G\"ottingen Graduate School for Neurosciences and Molecular Biosciences
(DFG Grant GSC 226/2) to B.S., the EPSRC via the grant EP/N013492/1
to C.B., the JST CREST, Grant Numbers JPMJCR14D2, JPMJCR15K1 to K.A.
and the Max Planck Society to M.T. 
\end{acknowledgments}

\subsection*{Author contributions}

B.S., D.W. and M.T. conceived and designed the research. B.S. acquired
the data, performed the data analysis and formulated stochastic predictions.
All authors contributed to discussing the results and writing the
manuscript. 

\subsection*{Competing interests}

The authors declare no competing interests.
\end{document}